\documentstyle[preprint,aps]{revtex}

\title{Ferromagnetic phase transition in a Heisenberg fluid:
Monte Carlo simulations and Fisher corrections to scaling}

\author{I. M. Mryglod,$^{1,2}$, I. P. Omelyan,$^1$ and R. Folk$^2$}

\address{$^1$Institute for Condensed Matter Physics,
         1 Svientsitskii Str., UA-79011 Lviv, Ukraine}
\address{$^2$Institute for Theoretical Physics, University of Linz,
         A-4040 Linz, Austria}

\date{\today}

\begin{document}

\maketitle

\begin{abstract}

The magnetic phase transition in a Heisenberg fluid is studied by
means of the finite size scaling (FSS) technique. We find that even
for larger systems, considered in an ensemble with fixed density, the
critical exponents show deviations from the expected lattice values
similar to those obtained previously. This puzzle is clarified by
proving the importance of the leading correction to the scaling that
appears due to Fisher renormalization with the critical exponent equal
to the absolute value of the specific heat exponent $\alpha$. The
appearance of such new corretions to scaling is a general feature of
systems with constraints. 

\vspace{4pt}
\noindent
Pacs numbers: 05.70.Jk; 75.40.-s; 75.50.Mm; 64.70.-p

\end{abstract}

\newpage
Monte Carlo (MC) simulations of finite systems near phase transitions
have received considerable attention in recent years \cite{Bin}. Of
notable current interest are continuum spin fluid models \cite{Set1}
which are considered as a first step towards the modelling of ferrofluids
\cite{Feld} and adsorption surface phenomena \cite{Sme}. Several
important results, which have both theoretical and experimental interest,
were obtained for such models. For example, it was found that the phase
diagrams for spin fluids due to the interplay between spin and translational
degrees of freedom are much more complicated \cite{Set1} than in
non-magnetic liquids. Magnetic ordered phases can exist both in gas and
liquid states. By applying an external magnetic field, one can shift
significantly the locus of the gas-liquid transition \cite{Set2} and change
the dynamic properties \cite{Set3}; both the static and dynamic properties
in the models discussed show differences from the non magnetic fluid and the
magnetic lattice model.

One important question in this context is whether the magnetic transition
in a Heisenberg fluid belongs to the same universality class as the
corresponding transition in the lattice model. On general grounds
(annealed systems) \cite{Fisher}, the lattice universality class is
expected. In Ref.~\cite{Nijmei1} using MC method, a novel set of critical
exponents was found that were in disagreement with the expected results.
Similar disagreements were later obtained for two- and three-dimensional
(2d and 3d) Ising fluids \cite{Nijmei2,Ferreira}, where Fisher renormalized
exponents were expected \cite{Fisher}. In all the cases mentioned, a weak
{\it dependence} of universal quantities on the density of particles $n=N/V$
and {\it systematic deviation} from the predicted critical exponents were
observed in the MC simulations. The general conclusion was that computer
simulations were strongly affected by {\it nonlinear crossover} effects,
which hide the true asymptotic critical behavior, giving only effective
critical exponents.

The goal of the present study is to resolve this puzzle by  performing a new
series of MC simulations for a Heisenberg fluid, considering larger finite
size systems, and to compare the results obtained with the previous data
\cite{Nijmei1}. Contrary to the belief that the true asymptotic critical
behavior would be observed, deviations from the expected exponents remained.
It is the aim of this Letter to show that a new correction term \cite{Fisher}
(Fisher correction term) has to be taken into account in the FSS even for very
large systems.

Let us consider a classical system, composed of $N$ magnetic particles
of mass $m$ and described by the Hamiltonian \cite{Set3,Nijmei1}
\begin{equation}  \label{H}
H = \sum_{i=1}^N \frac{m {{\bf v}_i}^2}{2} + \sum_{i < j}^N
\Big[ \ \Phi(r_{ij}) - J(r_{ij}) \, {\bf s}_i {\mbox{\boldmath
$\cdot$}} {\bf s}_j \Big],
\end{equation}
where ${\bf r}_i$ and ${\bf v}_i$ denote the position and velocity of
particle $i$ carrying spin ${\bf s}_i$. In our MC study, the liquid
subsystem potential $\Phi(r_{ij})$ was chosen to be of soft-core-like form,
$\Phi(r)=4\varepsilon [(\sigma/r)^{12}-(\sigma/r)^6] +\varepsilon$ at
$r<2^{1/6}\sigma$ and $\Phi(r)=0$ otherwise, and the exchange integral
$J(r_{ij})>0$, describing spin interactions, was modelled by the Yukawa
function, $J(r)= \epsilon(\sigma/r) \exp[(\sigma-r)/\sigma]$. The function
$J(r)$ was truncated at $R=2.5 \sigma$ and shifted to  zero at the truncation
point (this avoids force singularities during  MD calculations \cite{OMF}).
Staying within the classical approach we consider ${\bf s}_i$ as a three-component
continuous vector with a fixed length $|{\bf s}_i| = 1$.

The simulations were carried out in the basic cubic box $V=L^3$ (employing
periodic boundary conditions) at the reduced density of $n^\ast=N\sigma^3/V
=0.6$ for a reduced core intensity of $\varepsilon/\epsilon=1$. The number
of particles $N$ were taken as $N=$125, 256, 512, 1000, 2048, 4000,
8000, and 16384. The simulations have been
performed for five values of reduced temperature $T^\ast$, $T^\ast=
k_{\rm B} T/\epsilon=$2.000, 2.025, 2.050, 2.075, and 2.100. The system was
allowed to achieve equilibrium for 100 000 $N$ attempted moves. The total
number of trial moves per particle (cycles) performed in the equilibrium state
was $1\,000\,000$. The canonical averaging over the system was carried out using
a biasing scheme \cite{Craknell,Frenkel} for sampling orientational degrees
of freedom. To minimize computational costs, the cell list technique
\cite{Frenkel} was employed in handling the interparticle interactions.

The critical properties of a system in the thermodynamic limit may be extracted
from the behavior of finite size systems by examining the size dependence of
thermodynamic quantities \cite{Bin,Fish}. According to the FSS theory, various
thermodynamic quantities can be written in a scaling form $q(L,T) = L^{x_q/\nu}
{\cal Q} (z)$, where $L$ is a linear length of system, $x_q$ is a critical
exponent of the quantity $q$, and $z=tL^{1/\nu}$ is the temperature scaling
variable with $t= (T-T_c)/T_c$ ($T_c$ and $\nu$ are the bulk critical temperature
and critical exponent of correlation length $\xi$). Because we are interested
in zero-field properties, only the scaling variable $z$ appears in the scaling
function ${\cal Q}$.

There are several methods to determine the critical temperature $T_c$. One of
them is the Binder crossing technique \cite{Bin} formulated for the fourth-order
cumulant $U_4=1 - M_4/(3 M_2^2)$, where $M_l =\langle m^l \rangle$, $m =
|{\bf m}|$, and ${\bf m}= {N}^{-1} \sum_{i=1}^{N} {\bf s}_i$. This method does
not need any assumptions about critical exponents, but for small systems the
position of intersection points between any two curves $U_4$, related to systems
with lengths $L$ and $L'$, depends usually on $L$ and $L'$, because of corrections
to FSS. We have estimated the value of $T_c$ as the average over the cross-section
temperatures $T_{\rm cross}(L,L')$, found for systems with $L=L_i$ and $L'=L_{i+1}$
($i \geq 3$), giving $T_c = 2.055 \pm 0.001$. Here and below the increasing
subscript $i$ denotes increasing numbers of particles $N_i$ from the set $\{N_i\}$
considered. In a similar way, we found  $U_4^\ast=0.618 \pm 0.003$. The same
estimate for $T_c$ has been found by using the crossing technique for the
function $\xi (L_i,T)/L_i$  (see, e.g., \cite{Parisi}) within the phenomenological
renormalization group scheme. Using a more precise method for extracting $T_c$
\cite{Bin} from the values $T_{\rm cross}(L,L')$, obtained for the Binder parameter
$U_4$ with the fixed ratio $L_4/L_1=L_5/L_2=L_7/L_4= L_8/L_5=2$, we have obtained
$T_c = 2.054 \pm 0.001$ and $U_4^\ast=0.619 \pm 0.002$.

An alternative method, proposed in \cite{Landau2}, allows one to estimate
simultaneously both the critical temperature $T_c$ and the critical exponent $\nu$
within the same series of calculations. The main idea of this approach (the
scanning technique) is to look for a quantity-independent slope of the set of
functions $V_l$ with $l=1,2,\ldots,6$, all of which have similar scaling
behavior. These functions are defined via the derivatives $K_l= \partial M_l/
\partial \beta$ (see, for details, Ref.~\cite{Landau2}). The results are shown in
Fig.~1, so that we have got $T_c=2.057 \pm 0.001$ and $1/\nu=1.396 \pm 0.006$. Note
that within the scanning technique, no corrections to scaling have been taken into
account.

We have also used other known methods \cite{Bin} to estimate $T_c$. One of these
(the shifting technique) is based on the analysis of the size-dependent shift of
a peak $T_{\rm peak}(L)$, observed in some thermodynamic quantities (e.g.,
specific heat $C_V$, susceptibility $\chi$, derivatives $K_l$). If corrections to
scaling are neglected, the location of the peak $T_{\rm peak}(L)$ has the general
form $T_{\rm peak}(L) = T_c + A L^{-1/\nu}$, where $A$ is a quantity-dependent
constant. Note that in order to determine $T_c$ one has to estimate accurately
the exponent $\nu$ as well as the values $T_{\rm peak}(L)$. In the temperature
range considered, well-defined peaks for all the sizes $L_i$ have been observed
for the functions $\partial M / \partial \beta$, $\bar{U}_3 = (M_3-3M_2M_1+M_1^3)
/[M_1 (M_2-M_1^2)]$, and $\chi_3=N^2(M_3-3M_2M_1 +M_1^3)$ \cite{Landau2,Panag}.
For all these cases, our estimate of $T_c\simeq 2.058 \pm 0.002$, found with
$1/\nu=1.396$ for the five largest sizes $L$, is in agreement with the 
scanning technique but not with the crossing technique. Moreover, the dependence
$T_{\rm peak}(L)$ versus $L^{-1/\nu}$ showed a pronounced curvature for smaller
system sizes $L$. Hence, {\it the first puzzle uncovered in our study is the
disparity in the estimates for $T_c$} found using two types of standard FSS
techniques, namely, (i) the crossing technique for Binder parameter as well as
for the correlation length, and (ii) the scanning and shifting techniques. This
disparity could not be entirely explained by the error bars and indicates strong
crossover effects. Note that corrections to scaling were completely neglected in
the methods of type (ii). In order to clarify the cause of the difference found
for $T_c$, the scanning technique was used for the four largest systems only,
giving $T_c=2.054 \pm 0.001$ in agreement with the result of the crossing method.
However, despite the {\it improvement} in $T_c$, the value $1/\nu=1.312 \pm 0.007$
was a {\it deterioration}. This is {\it another puzzle} which needs explanation.
Taking everything together these results support the presence of strong crossover
effects in the system considered.

Knowing $T_c$, one can then estimate the critical exponents again using the FSS
theory \cite{Bin}. We have calculated the exponent ratios $\beta/\nu$ and
$\gamma/\nu$ the FSS behavior of $M(L,T_c)$ and the magnetic susceptibility
$\chi(L,T_c)$. For $T_c= 2.057$, we found  $\beta/\nu=0.544 \pm 0.015$ and
$\gamma/\nu= 1.90 \pm 0.03$, respectively. Nearly the same estimates were
obtained using other FSS methods. Taking $T_c=2.054$ for the four largest sizes
$L_i$, we obtained $\beta/\nu=0.520 \pm 0.008$ and $\gamma/\nu= 1.87 \pm 0.03$.
The results for both choices $T_c=2.057$ and $T_c=2.054$ are summarized in
Table~1 in the first and second lines, respectively, beneath the Fluid$^2$
heading. The estimates for $1/\nu$ were found by the scanning technique.

Comparing our results with the previous ones \cite{Nijmei1}, we conclude that:
(i) the ratios of critical exponents $\beta/\nu$ and $\gamma/\nu$ found in our
study are closer to the values known for the lattice model \cite{Landau2,Holm};
(ii) the critical exponent $\nu$ is extremely sensitive to the estimate of
critical temperature $T_c$ used; (iii) even for larger systems, which this
study considers, a systematic deviation from the lattice exponents is seen
that cannot entirely be justified by the error bars; and (iv) the disparity in
estimates found for the critical temperature $T_c$, using the crossing and
shifting techniques, has no explanation within the standard FSS approach.
Hence, our data have to be considered as results for {\it effective exponents}
and one can expect that the true asymptotic behavior would be visible only for
much larger systems. If non-asymptotic crossover effects are considered, one may
think of the presence of the Wegner correction term. However, this is expected
to be negligible for our largest system sizes. One has to ask, therefore, what
the reason is for such a strong crossover in the system we considered, 
compared to the lattice model.

In order to investigate this problem in more detail let us recall an idea
encountered in the Fisher renormalization \cite{Fisher} for a system under
thermodynamic constraint. According to this idea, the critical singularities
in the grand canonical ensemble, with fixed chemical potential $\mu$, may be
different from those describing the system in the canonical ensemble with fixed
density $n$. One has to performed the corresponding Legendre transformation
carefully, taking into account the properties of singular functions in the grand
canonical ensemble. In particular, this gives the well-known relation (see, e.g.,
Eq.~(2.38a) in \cite{Fisher})
\begin{equation} \label{main}
\tau = a_0 t^{x_{\alpha}} \Big(1+a_1 t^{\Delta_{\alpha}} \Big),
\end{equation}
which connects the reduced temperature scales $t$ and $\tau$ in the two
different ensembles with fixed $n$ and $\mu$, respectively. The values of
$x_{\alpha}$ and $\Delta_{\alpha}$ in (\ref{main}) depend on the sign
of the specific heat critical exponent $\alpha$, and are equal to
$(1-\alpha)^{-1}$ or 1, and $\alpha(1-\alpha)^{-1}$ or $-\alpha$ for $\alpha$
positive or negative, respectively. It is seen already from (\ref{main}) that
{\it independent of whether Fisher renormalization changes the critical
exponents in the ensemble with fixed $n$, a new type of corrections to scaling
appears in the canonical ensemble}. These corrections, being proportional to
$\alpha$, must not be confused with Wegner corrections to scaling; and because
of the smallness of $\alpha$ in the Heisenberg universality class, they have to
be taken into account within the FSS analysis. We note also that Eq.~(\ref{main})
is not the only source \cite{MOF} for the appearance of new corrections (as was
assumed, e.g., in \cite{Krech}). There is another reason, which also follows from
thermodynamics. For example, using hyperscaling relations for the critical
exponents, it can be easily proved \cite{MOF} that the second term in the
known expression
$
{\chi}_{T,n} = \left({\partial M}/{\partial h}\right)_{T,n} =
\left({\partial M}/{\partial h}\right)_{\mu} -
({\partial M}/{\partial \mu})_{T}
\left({\partial N}/{\partial \mu} \right)^{-1}_{T}
({\partial N}/{\partial h})_{\mu}
$
produces an additional correction to the magnetic susceptibility $\chi_{T,n}$
with an exponent proportional to $\alpha$. Hence, these new corrections to scaling
cannot be included in the standard FSS by means of simple rescaling of $t$, which
follows from (\ref{main}) and as was proposed in \cite{Krech}.

In order to prove our predictions and to estimate the range of asymptotic
behavior in which the new correction can be neglected, we have performed additional
calculations. In Fig.~2 the results, obtained for the temperatures $T_{\rm peak}(L)$,
where the maximums of the functions $\partial M / \partial \beta$ and ${\chi}(T,L)$
are located, are shown for different sizes of $L$. These results have been fitted
(dashed lines) for the {\it four largest} system sizes by using the expression
\begin{equation}    \label{Bin_2}
T_{\rm peak}(L) = T_c + A N^{-1/3\nu}
\left( 1+ B N^{-|\alpha|/3\nu}\right),
\end{equation}
with the values of $1/\nu$ and $\alpha$ known for the lattice model
\cite{Landau2,Holm}. The quantity-dependent constants $A$ and $B$ were then
estimated. It is seen in Fig.~2 that: (i) the fitting curves are
in rather good agreement with the MC data obtained even for smaller values of $L$;
(ii) such a simple procedure allows one to understand the strong deviation from
the linear dependence $T_{\rm peak}(L) = T_c + A N^{-1/3\nu}$ that follows from
(\ref{Bin_2}) when the correction to scaling is neglected; and (iii) the disparity
in estimates found for $T_c$ within the crossing and shifting techniques can be
explained. Using the fitting procedure, described above, we have found that the
estimate $T_c= 2.055 \pm 0.001$ gives a rather good fit for all the data
$T_{\rm peak}(L)$, obtained from the maximum positions of $\partial M / \partial
\beta$, ${\chi}(T,L)$, $\bar{U}_3 (T,L)$, and ${\chi}_3(T,L)$. Another finding
was that in contrast to the strong quantity-dependence of $A$, the parameter $B$
in (\ref{Bin_2}) is almost independent of the quantity considered \cite{xi}. Note that if
the rescaling relation (\ref{main}) is considered as the unique reason for the
appearance of the new correction, then the parameter $B$ is quantity-independent.
From the fitting procedure it has been found that $B \simeq 1.3 \pm 0.2$ for all
the five sets of $T_{\rm peak}(L)$ studied. Having the value of $B$, we can then
estimate the minimal number of particles $N_{\rm min}$ such that the correction
term in (\ref{Bin_2}) can be neglected if $N> N_{\rm min}$. This gives $N_{\rm min}
\simeq 10^8$ (then the relative contribution of the second term in the bracket
of (\ref{Bin_2}) is less than 0.5), and, therefore, it is clear why the true
asymptotic behavior could not be observed earlier \cite{Nijmei1}, or in  
our MC study. Finite systems with $N> N_{\rm min}$ have  not been considered
so far in MC simulations. Hence, only the effective exponents could be studied
for smaller size systems.

In conclusion, we note that if the absolute specific heat exponent $\alpha$ is
small enough, Fisher corrections to scaling discussed are very important in
models {\it with constraints}. This holds at $d=3$ for the Ising, the XY and the
Heisenberg classes of universality, as well as for other systems. In particular,
we are convinced that the problems found in the Ising fluid \cite{Nijmei2,Ferreira}
have the same origin. In this respect it is also worth referring the reader to
Refs.~\cite{corr}, where, within the $\epsilon$-expansion scheme, it was proven
analytically that the leading correction to scaling in a compressible Heisenberg
magnet as well as in a randomly diluted, weakly inhomogeneous Heisenberg model
is equal to $-\alpha$, and this supports our conclusions. More detailed
results including the determination of the values of the asymptotic exponents
will be given elsewhere.

We thank M.Fisher and M.Anisimov for their interest and their suggestions.
Part of this work was supported by the Fonds zur F\"orderung der
wissenschaftlichen Forschung under Project No. P12422-TPH.

\newpage
{
TABLE 1. Summary of results with $n$ being the reduced density. In the first
three rows (denoted as Fluid$^1$) the results from \cite{Nijmei1} are gives.
In the last row the universal quantities known for the lattice Heisenberg model
\cite{Landau2,Holm} are presented. Our results are shown in the rows denoted as
Fluid$^2$.}

\vspace{1.5cm}
\begin{center}
\begin{tabular}{c|c|c|c|c} \hline\hline
      &  $U_4$  & $1/\nu$ & $\beta/\nu$ & $\gamma/\nu$ \\  \hline
\ Fluid$^1$ \ &         &           &           &  \\
\ \ n=0.4 \ & \ 0.613 \ & \ 1.35(5) \ & \ 0.55(2) \ & \ 1.86(3) \\
\ \ n=0.6 &  0.608  &  1.41(3)  &  0.56(2)   & 1.85(1) \\
\ \ n=0.7 &  0.605  &  1.42(3)  &  0.55(2)   & 1.84(3) \\ \hline

\ Fluid$^2$ \ &     &           &            &         \\
\ \ n=0.6 \ &  0.619  &  1.40(1)  &  0.54(2) &   1.90(3) \\
\ \         &         &  1.31(1)  &  0.52(1) &   1.87(3) \\
 \hline
\ Lattice &  0.622  &  1.421(5) &  0.514(1)  & 1.973(2) \\
\hline\hline
\end{tabular}
\end{center}

\newpage
\begin{center}
{\bf \Large FIGURE CAPTIONS}
\end{center}

\vspace{1.5cm}
{FIG. 1. Quantity dependence of scanning results for the functions $V_l$
possessing the same scaling properties \cite{Landau2}. The horizontal line
for $T_c=2.057$ is drawn at $1/\nu=1.396$.}

\vspace{1.5cm}
{FIG. 2. Size dependence of the maximum locations in the derivative
$\partial M / \partial \beta$ (triangles) and susceptibility $\chi(T,L)$
(diamonds). The results of fitting to MC data (dashed curves) are found
using (\ref{Bin_2}) with $1/\nu$ and $\alpha$ known for the lattice model.}


\begin{thebibliography}{19}
\bibitem{xi} This results from the fact that the scaling of $L$ goes with
the correlation length and therefore the expression in the bracket of
Eq.(\ref{Bin_2}) is expected to be the same in all cases.
\end{thebibliography}
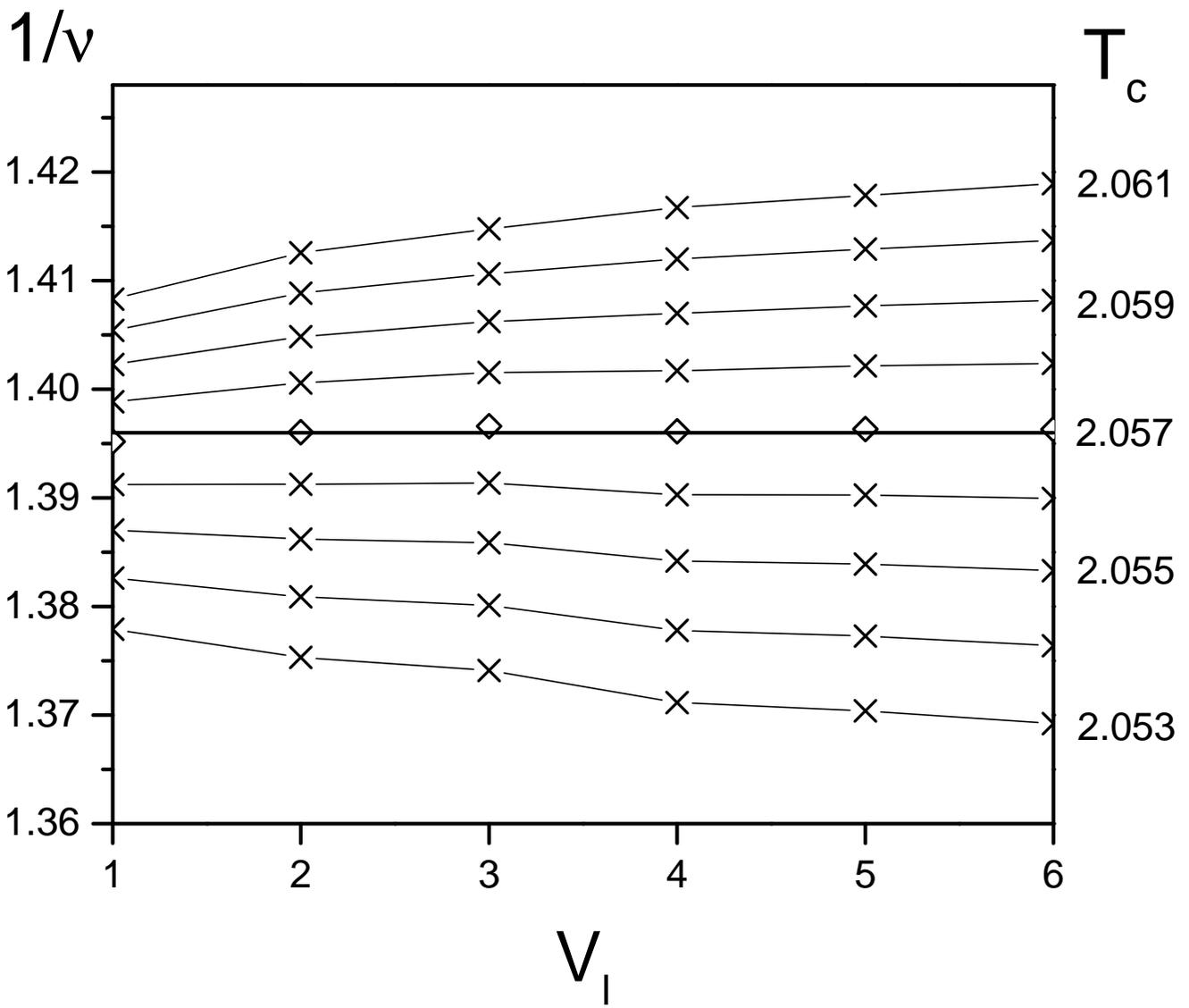
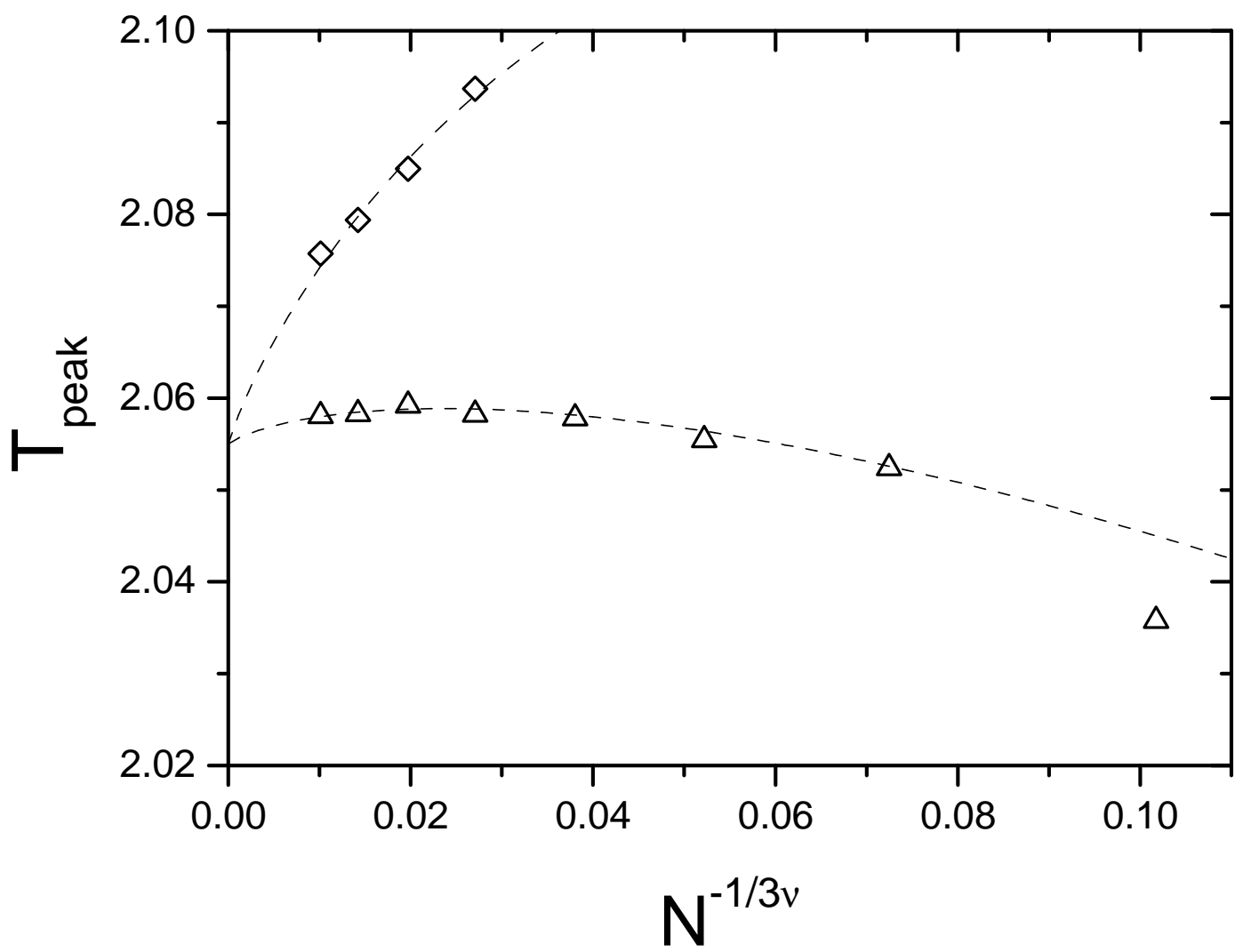
\end{document}